\documentclass[preprint2, longabstract]{aastex}

\begin{document}


\title{High-resolution IR absorption spectroscopy of polycyclic aromatic hydrocarbons in the 3-$\mu$m region: Role of periphery}

\author{Elena Maltseva}
\affil{University of Amsterdam, Science Park 904, 1098 XH Amsterdam, The Netherlands}
\and
\author{Annemieke Petrignani\altaffilmark{1,2}, Alessandra Candian, Cameron J. Mackie}
\affil{Leiden Observatory, Niels Bohrweg 2, 2333 CA Leiden, The Netherlands}
\altaffiltext{1}{Radboud University, Toernooiveld 7, 6525 ED Nijmegen, The Netherlands}
\altaffiltext{2}{University of Amsterdam, Science Park 904, 1098 XH Amsterdam, The Netherlands}
\and
\author{Xinchuan Huang\altaffilmark{3}}
\affil{SETI Institute, 189 Bernardo Avenue, Suite 100, Mountain View, CA 94043, USA}
\altaffiltext{3}{NASA Ames Research Center, Moffett Field, California 94035-1000, USA}
\and
\author{Timothy J. Lee}
\affil{NASA Ames Research Center, Moffett Field, California 94035-1000, USA}
\and
\author{Alexander G. G. M. Tielens}
\affil{Leiden Observatory, Niels Bohrweg 2, 2333 CA Leiden, The Netherlands}
\and
\author{Jos Oomens}
\affil{Radboud University, Toernooiveld 7, 6525 ED Nijmegen, The Netherlands}
\and
\author{Wybren Jan Buma}
\affil{University of Amsterdam, Science Park 904, 1098 XH Amsterdam, The Netherlands}
\email{w.j.buma@uva.nl}
%
%
%

\begin{abstract}
In this work we report on high-resolution IR absorption studies that provide a detailed view on how the peripheral structure of irregular polycyclic aromatic hydrocarbons (PAHs) affects the shape and position of their 3-$\mu$m absorption band. To this purpose we present mass-selected, high-resolution absorption spectra of cold and isolated phenanthrene, pyrene, benz[a]antracene, chrysene, triphenylene, and perylene molecules in the 2950-3150 cm$^{-1}$ range. The experimental spectra are compared with standard harmonic calculations, and anharmonic calculations using a modified version of the SPECTRO program that incorporates a Fermi resonance treatment utilizing intensity redistribution. We show that the 3-$\mu$m region is dominated by the effects of anharmonicity, resulting in many more bands than would have been expected in a purely harmonic approximation. Importantly, we find that anharmonic spectra as calculated by SPECTRO are in good agreement with the experimental spectra. Together with previously reported high-resolution spectra of linear acenes, the present spectra provide us with an extensive dataset of spectra of PAHs with a varying number of aromatic rings, with geometries that range from open to highly-condensed structures, and featuring CH groups in all possible edge configurations. We discuss the astrophysical implications of the comparison of these spectra on the interpretation of the appearance of the aromatic infrared 3-$\mu$m band, and on features such as the two-component emission character of this band and the 3-$\mu$m emission plateau.
\end{abstract}




\section{Introduction}

Polycyclic Aromatic Hydrocarbons (PAHs) are a family of molecules consisting of carbon and hydrogen atoms combined into fused benzenoid rings.  From a chemical and physical point of view they have properties that have led  to exciting applications in novel materials \citep{Wan2012, Sullivan2008}, but at the same time also to quite a cautious use on account of their health-related impact \citep{Kim2013, Boffetta}. For astrophysics they play a particularly important role since PAHs have been proposed as main candidates for carriers of the so-called aromatic infrared bands (AIBs), a series of infrared emission features that are ubiquitously observed across a wide variety of interstellar objects. These emission features are thought to be nonthermal in nature and arising from radiative cooling of isolated PAHs that have been excited by UV radiation \citep{Sellgren1984}.

Since they offer such a powerful probe for carbon evolution in space, these bands have been subject to extensive experimental and theoretical research with the ultimate aim being a rigorous identification of the molecular structure of the AIB carriers. Significant progress has been made in this respect with infrared (IR) studies on PAH species deposited in a cold (10 K) rare-gas matrix (for example, \citet{Hudgins1998,Hudgins1998b,Hudgins2000}. Although such cooling conditions allow for an increase in spectral resolution as compared to room-temperature experiments, they lead at the same time to matrix-induced effects that are not well understood and hard to predict. Gas-phase studies are much preferred but have so far predominantly been restricted to IR absorption studies of hot (1000 K) vaporized PAHs \citep{Joblin1995a,Joblin1994b} or at best under room-temperature conditions (e.g., \citet{Pirali2009,Cane1997}). Due to their low volatility, high-resolution studies of low-temperature, isolated PAH molecules has for a long time remained out of reach with as a notable exception the cavity ring down spectroscopy (CRDS) studies  of \citet{Huneycutt2004} on small PAHs, although contaminations originating from isotopologues and other PAH species or impurities remained a point of concern.

Recently, we have applied IR-UV double resonance laser spectroscopic techniques on PAHs seeded in supersonic molecular beams. In combination with mass-resolved ion detection these techniques allow for recording of mass- and conformation-selected IR absorption spectra with resonance band widths down to 1 cm$^{-1}$ \citep{Maltseva2015}. Under such high-resolution conditions, IR absorption spectra of PAHs in the 3-$\mu$m region turn out to display an unexpected large number of strong bands, and certainly many more than expected on the basis of a simple harmonic vibrational analysis. Such a conclusion is more pertinent as theoretical studies of IR spectra of PAHs are typically performed at the Density Functional Theory (DFT) level, using the harmonic approximation for vibrational frequencies and double harmonic approximation for intensities, neglecting the effects of anharmonicity. In our previous papers we demonstrated that a proper treatment of anharmonicity and Fermi resonances indeed leads to predicted spectra that are in near-quantitative agreement with the experimental spectra, both with respect to the frequencies of vibrational bands and their intensities.

The shape of the 3-$\mu$m band recorded in astronomical observations has been found to vary within the same astronomical object and between different astronomical objects. To account for these differences, several explanations have been put forward \citep{Sellgren1990, Tokunaga1991, VanDiedenhoven2004}. One of the suggestions for classifying the shape of this band is to interpret it as being associated with emission from two components \citep{Song2003a, Candian2012} in which there are contributions from different groups of carriers at 3.28 and 3.30 $\mu$m. \citet{Song2007} propose that these two components originate from groups of PAHs with different sizes, finding support for this in the laboratory high-temperature gas-phase studies of pyrene ($C_{16}H_{10}$) and ovalene ($C_{32}H_{14}$) for which a blue shift of the 3-$\mu$m band was observed upon increasing the size of the PAH \citep{Joblin1995a}. Another factor that has been suggested to contribute to the apparent two-component appearance of the emission are differences in the peripheral structure of different PAHs, in particular steric effects as occurring for hydrogen atoms at so-called bay-sites \citep{Candian2012}. The influence of the edge structure has been investigated by means of harmonic DFT calculations for large species \citep{Bauschlicher2009,BauschlicherCharlesW2008} but systematic experimental high-resolution studies are notoriously lacking. Similarly, it has been found \citep{Geballe1989} that next to the prominent emission band at 3.29 $\mu$m, a broad plateau is present that spans the 3.1-3.7 $\mu$m region and has been indicated as the 3-$\mu$m plateau. It has been speculated that this plateau might in part derive from anharmonic couplings to vibrational combination levels \citep{Allamandola1989a}. However, to what extent this explanation can account for the appearance of the entire plateau is still far from clear.

Our previous study aimed at recording spectra under the highest resolution conditions possible and applying the appropriate theoretical treatment including anharmonic effects and resonances. For that reason, we focused on the spectra of the linear PAHs naphthalene (C$_{10}$H$_{8}$), anthracene (C$_{14}$H$_{10}$), and tetracene (C$_{18}$H$_{12}$). As discussed above, astronomical spectra likely comprise the contributions of a much larger variety of PAHs. To advance the interpretation and characterization of these data, we therefore extend our experimental and theoretical studies to a wider variety of condensed and irregular isomers containing up to five rings (phenanthrene C$_{14}$H$_{10}$, benz[a]antracene C$_{18}$H$_{12}$, chrysene C$_{18}$H$_{12}$, triphenylene C$_{18}$H$_{12}$, pyrene C$_{16}$H$_{10}$ and perylene C$_{20}$H$_{12}$). The goals of these studies are twofold. Firstly, we aim to understand how the effects of anharmonicity observed for the linear PAHs are affected by geometrical structure and how this in turn affects the appearance of the IR absorption spectra in the 3-$\mu$m region. Secondly, we aim to uncover general trends in band shapes that could provide spectral signatures that would allow for a much more detailed description of the contribution of different PAHs.

\section{Methods}
\subsection{Experimental techniques}
IR spectra of cold and isolated molecules were obtained in molecular beam setup described in \citet{Smolarek2011}. In these experiments the sample of interest was placed in an oven attached to a pulsed valve (General Valve). Sufficient vapor pressure was found to be obtained when the sample was heated to temperatures slightly higher than its melting point. Subsequent pulsed expansion with 2 bars of argon as carrier gas and a typical opening time of 200 $\mu$s then led to supersonic cooling of the sample molecules.

The ground-state vibrational manifold of PAHs was probed by UV-IR ion dip spectroscopy. To this purpose two-color resonance enhanced multiphoton ionization followed by mass-selected ion detection was used to generate an ion signal. In these experiments PAHs were electronically excited by fixing a frequency-doubled pulsed dye laser (Sirah Cobra Stretch) on a) the  $S_{1}$ ($^{1}A_{1}$) $\leftarrow$  $S_{0}$ ($^{1}A_{1}$) 0 - 0 transition at 29329.8 cm$^{-1}$ for phenanthrene \citep{Amirav1984b}; b) the $S_{1}$ ($^{1}A^\prime$) $\leftarrow$ $S_{0}$ ($^{1}A^\prime$) 0 - 0 transition at 26534.4 cm$^{-1}$ for benz[a]anthracene \citep{Wick1993};  
c) the  $S_{1}$ ($^{1}B_{u}$) $\leftarrow$  $S_{0}$ ($^{1}A_{g}$) 0 - 0 transition at 28194.3 cm$^{-1}$ for chrysene \citep{Zhang2012}; 
d) the $S_{1}$ ($^{1}B_{2u}$) $\leftarrow$  $S_{0}$ ($^{1}A_{g}$) 0 - 0 transition at 27210.9 cm$^{-1}$ for pyrene (which is close to but not exactly the same as the previously reported value by \citet{Ohta1987};  
e) the $S_{1}$ ($^{1}A_{1}^{\prime}$) $\leftarrow$ $S_{0}$ ($^{1}A_{1}^{\prime}$) transition to the 249 (29867) cm$^{-1}$ (1e′) level in the excited state of triphenylene \citep{Harthcock2014, Kokkin2007}; and f) the $S_{1}$ ($^{1}B_{3u}$) $\leftarrow$ $S_{0}$ ($^{1}A_{g}$) 0 - 0 transition at 24067.4 cm$^{-1}$  for perylene \citep{deVries1991}. Molecules in electronically excited states were ionized by an ArF excimer laser (Neweks PSX-501) in temporal overlap with the electronic excitation laser.

Prior to the pump and ionization laser beams needed to generate this signal, an IR laser beam with a linewidth of 0.07 cm$^{-1}$ and a typical pulse energy 1 mJ was introduced with a time delay of 200 ns. The 3 $\mu$m beam was generated by difference frequency mixing of the fundamental output of a dye laser (Sirah Precision Scan with LDS798 dye) and the 1064 nm fundamental of Nd:YAG laser (Spectra Physics Lab 190) in a LiNbO$_{3}$ crystal. Resonant excitation of vibrational levels was observed by a decrease of the ion signal due to depletion of the ground state population, allowing for recording of IR absorption spectra. Such IR spectra were recorded between 3.17 and 3.40 $\mu$m (2950 and 3150 cm$^{-1}$). With the present S/N ratios, no other IR bands could be observed outside this range.

\subsection{Computational methods}

Two types of theoretical IR spectra were produced: the standard harmonic vibrational approach implemented in Gaussian09 \citep{g09}, and an anharmonic vibrational approach that employs both Gaussian09 and a locally modified version of the program SPECTRO \citep{Mackie2015a,Gaw1996}, referred to in this work as G09-h and SP16 calculations respectively. Both G09-h and SP16 calculations start from DFT calculations that employ a similar integration grid as in \citet{Boese2004}, the B9-71 functional \citep{Hamprecht1998}, and the TZ2P basis set \citep{Dunning1971}, all of which has been found to give the best performance on organic molecules \citep{Boese2004,Cane2007}. The SP16 calculations utilize the quadratic, cubic, and quartic force constants calculated by Gaussian09 and transformed into Cartesian derivatives as the input for SPECTRO, which then performs its own vibrational second-order perturbation (VPT2) analysis \citep{Mackie2015,Mackie2015a}. SP16 treats the polyads of multiple simultaneous resonances (modes falling within 200 cm$^{-1}$ of each other) and allows for the redistribution of intensity among the resonant modes, which is a function not currently employed in Gaussian09. A more thorough account of the theoretical aspect of this work is given in a separate publication \citep{Mackie2016}. Harmonic calculations in this work were scaled in order to be compared with experimental data using a scaling factor (sf) of 0.961, while SP16 output does not require scaling and can be compared directly. \textbf{Both harmonic and anharmonic calculations are convolved with a 1 cm$^{-1}$ Gaussian line shape.}

\section{Results}

In the following we will discuss IR-UV ion dip spectra of phenanthrene, benz[a]antracene, chrysene, triphenylene, pyrene and perylene. In the first instance, we will present these spectra and compare them where possible with previous studies, highlighting only salient features and differences between the various PAHs. We will make extensive comparisons with the results of theoretically predicted spectra. A more detailed discussion of these calculations and the assignment of each of the observed bands in the experimental spectra will be presented in a separate study \citep{Mackie2016}. Subsequently, we will make a global comparison between the spectra, and discuss their implications for extrapolating to the 3-$\mu$m region of larger PAHs. Finally, we will consider the current interpretation of astronomical PAH data in the light of the present results.

\subsection{Phenanthrene}

\begin{figure}[]
\includegraphics[scale=0.5]{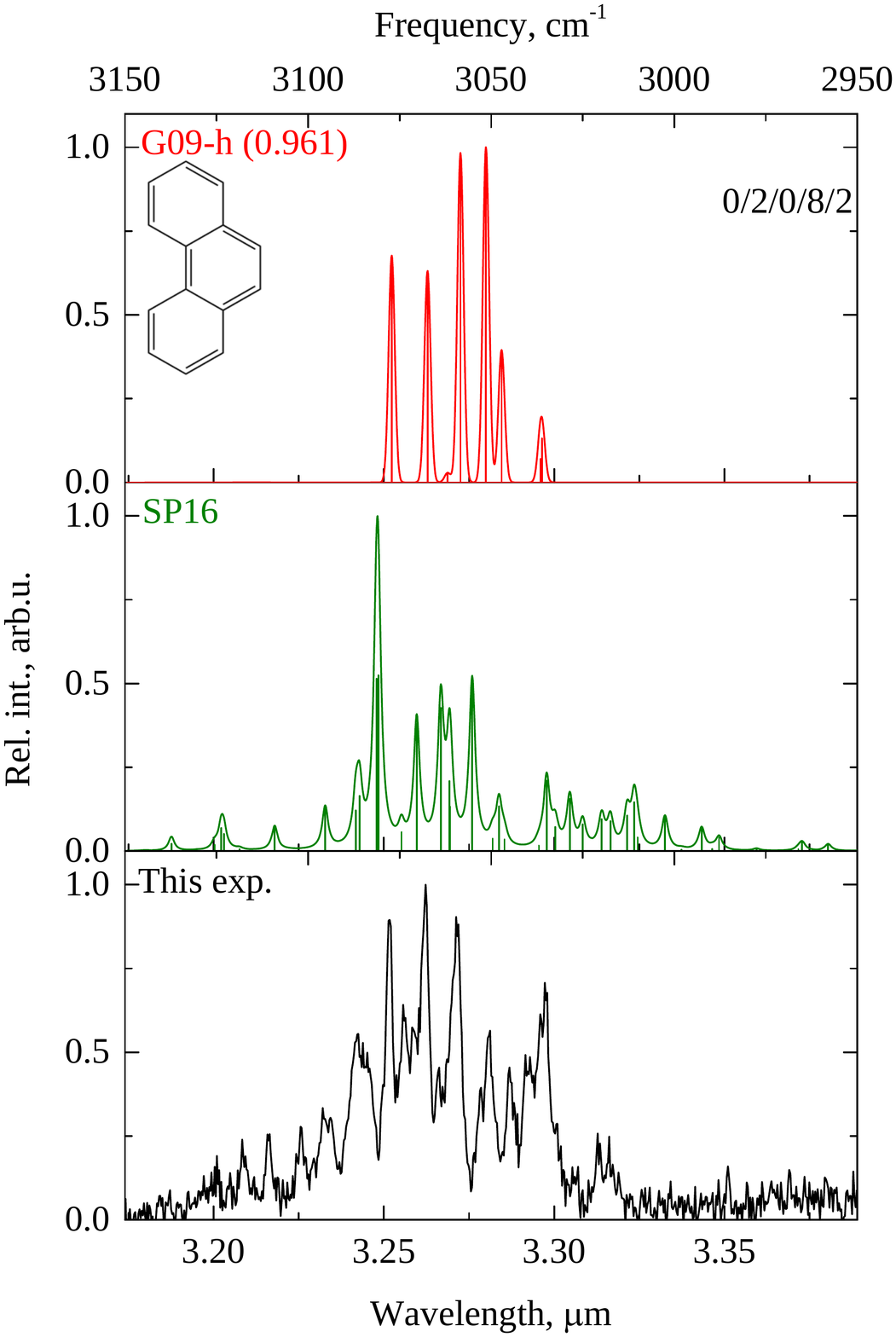}
\caption{The IR absorption spectrum of phenanthrene as predicted by G09-h (scaling factor sf= 0.961) and SP16 calculations (not scaled) together with the molecular beam gas-phase spectrum as measured in the present experiments. Number of hydrogens is mentioned in the following order: solo/duo/trio/quartet/bay H's. See Fig.\ref{Fig8bay} for more details. \label{Fig1Phen}}
\end{figure}

The 3-$\mu$m absorption spectrum of phenanthrene is shown in Figure \ref{Fig1Phen}, bottom panel. In total twenty-three experimental bands with a linewidth 2.5-4.5 cm$^{-1}$ are observed. The positions and strengths of these bands are reported in Table \ref{table1}. The most active region (where the bands are at least 40\% intensity of the dominant band) is located between 3030 and 3090 cm$^{-1}$ with the most  intense transition at 3065.3 cm$^{-1}$. This region is accompanied by two weak bands in the low-energy region and five overlapping weak bands in the high-energy wing. Table \ref{table1} also provides the values of bands reported previously in the CRDS studies of \citet{Huneycutt2004}. Due to better cooling conditions in the present experiments our absorption spectrum shows bands that are more narrow. As a result, nine additional bands can be discerned, although we do not find evidence for the weak band previously reported at 3025.3 cm$^{-1}$. Overall, the agreement between the two sets of data is good with the largest deviation for most of the bands not exceeding 2 cm$^{-1}$.

Phenanthrene belongs to the C$_{2\nu}$ point group. Within the harmonic approximation (Fig. \ref{Fig1Phen}, top panel) only eight modes are therefore IR active in the 3-$\mu$m region. The experimental spectrum, on the other hand, displays twenty-three bands and is a clear indicator of the failure of the harmonic approximation. We conclude that, similar to linear PAHs  \citep{Maltseva2015}, the CH-stretch region is dominated by Fermi resonances. Previously, we  showed \citep{Mackie2015} that in order to obtain good agreement between experimental and predicted spectra, VPT2 treatments are required that take intensity sharing and polyad resonances as implemented in SP16 into account. With such calculations, the redistribution of intensities over fundamental and combination bands -- in this case overtones are not accessible due to symmetry restrictions -- of the same symmetry leads to observable activity of many additional bands. Comparison between the 3-$\mu$m absorption spectrum of phenanthrene predicted by SP16 (Fig.\ref{Fig1Phen}, middle panel) with the experimentally obtained spectrum indeed shows good agreement.

\subsection{Pyrene}

\begin{figure}[]
\includegraphics[scale=0.5]{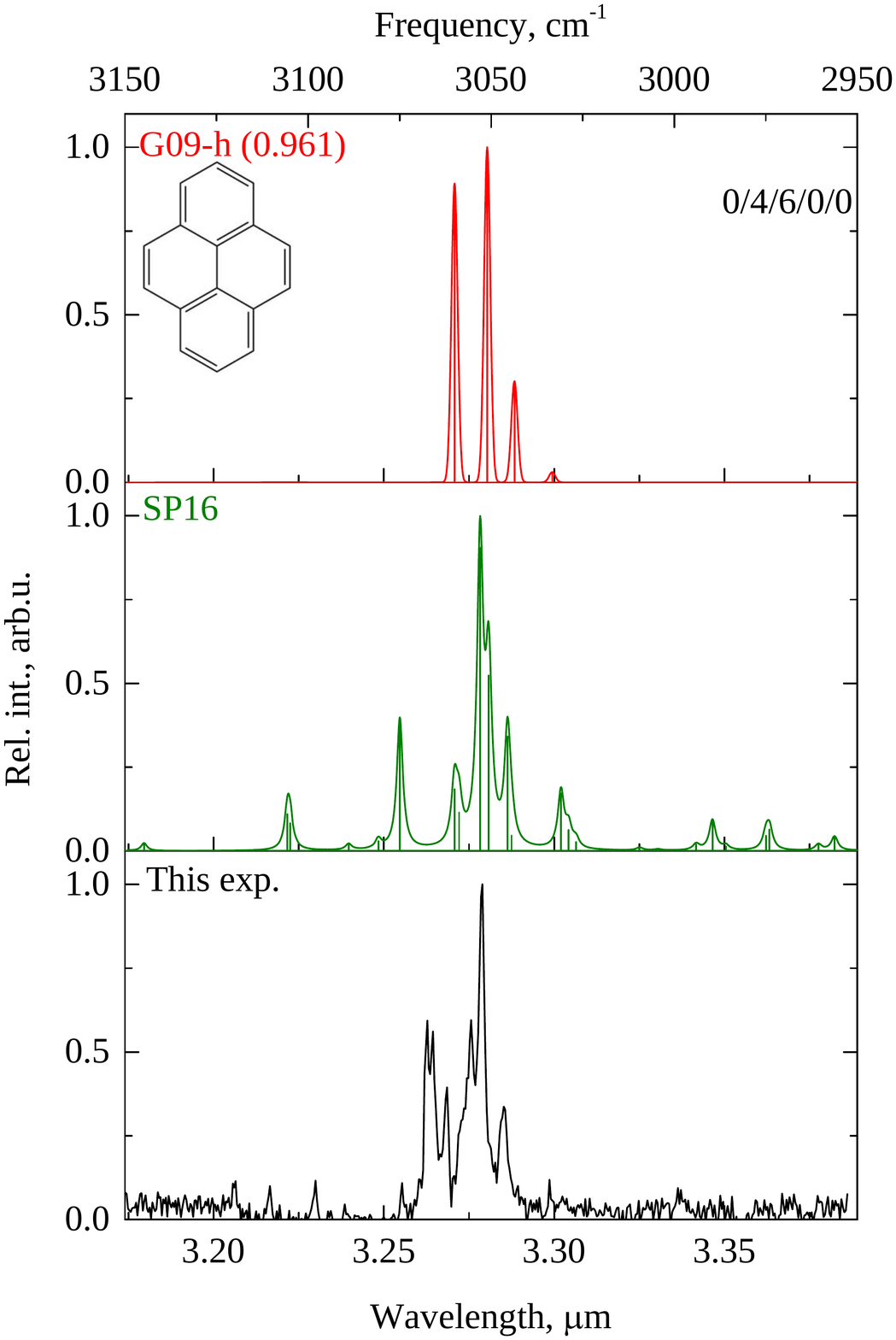}
\caption{The IR absorption spectrum of pyrene as predicted by G09-h and SP16 calculations together with the molecular beam gas-phase spectrum as measured in the present experiments. \label{Fig2Pyr}}
\end{figure}

Figure \ref{Fig2Pyr} (bottom panel) displays the IR absorption spectrum of pyrene in the 2950-3150 cm$^{-1}$ region as measured in the present study together with spectra predicted by G09-h and SP16 calculations. Compared to phenanthrene, the dominant vibrational activity is observed in a more compact region of 3045-3065 cm$^{-1}$ with the maximum intensity at 3049.8 cm$^{-1}$. The experimental spectrum shows fourteen relatively narrow bands with a linewidth of 1.7-3.5 cm$^{-1}$ whose positions are given in table \ref{table1}. This table also contains the positions of the ten bands reported previously by \citet{Huneycutt2004}, which agree well with the present observations (deviations less than 1 cm$^{-1}$). Our S/N ratio, on the other hand, allows us to identify weak bands at 3044, 3087.8, and 3096 cm$^{-1}$ that could previously not be discerned, but does not confirm the previously reported band at 3083.9 cm$^{-1}$. Due to the D$_{2h}$ symmetry of pyrene, only five IR active CH-stretch modes are expected in the harmonic approximation (Fig.\ref{Fig2Pyr}, top panel), which is clearly at odds with the experiment. Indeed, the polyad VPT2 anharmonic calculations (Fig.\ref{Fig2Pyr}, middle) are required to bring experiment and theory into better agreement.

\subsection{Benz[a]anthracene}

\begin{figure}[]
\includegraphics[scale=0.5]{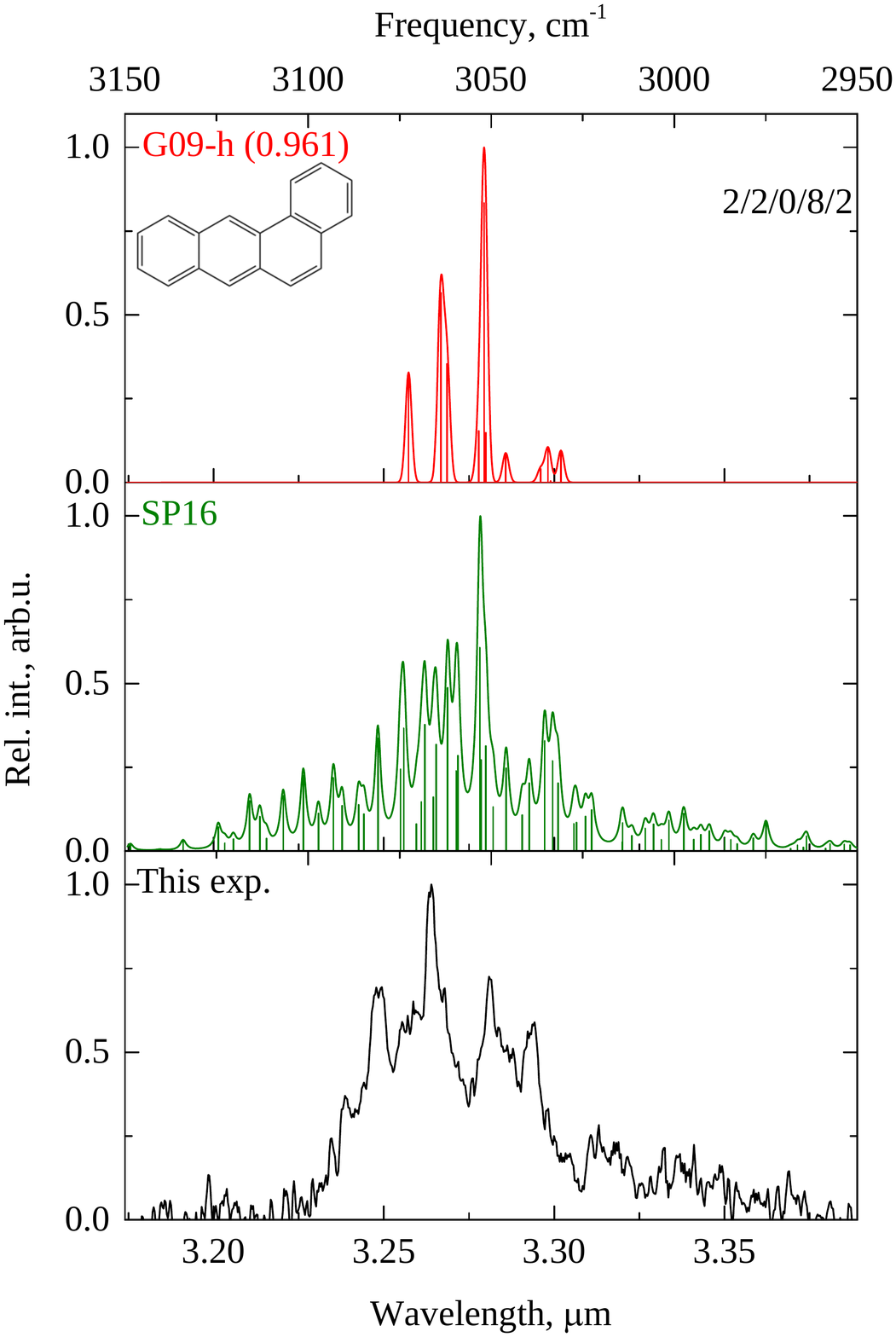}
\caption{The IR absorption spectrum of benz[a]anthracene as predicted by G09-h and SP16 calculations together with the molecular beam gas-phase spectrum as measured in the present experiments.\label{Fig3Benz}}
\end{figure}

The IR absorption spectrum benz[a]anthracene in the 2950-3150 cm$^{-1}$ region is shown in Figure \ref{Fig3Benz} together with the spectra predicted by harmonic and anharmonic calculations. The main spectrum feature covers quite a large frequency range and lacks the narrow bands observed for the other PAHs reported here. Instead a broad structure between 3030 to 3090 cm$^{-1}$ with the highest intensity at 3063.8 cm$^{-1}$ is seen. In this structure only five bands can be distinguished clearly due to the overlap of a large number of blended transitions in the present resolution. 

The presently obtained spectrum is the first high-resolution gas-phase IR spectrum that has been reported for benz[a]anthracene. We therefore compare in Table \ref{table1} the line positions as derived from Figure \ref{Fig3Benz} with the Matrix Isolation Spectroscopy (MIS) data of \citet{Hudgins1998b}. We find larger deviations of up to 3 cm$^{-1}$ which likely find their origin in matrix-induced perturbations, difficulties in determination of the center of bands due to their shape (as occurs for the most intense band), and the noise level (as occurs for the weak bands). Interestingly, our spectrum does not give an indication for the presence of a band at 3119.5 cm$^{-1}$ that was reported in the MIS studies, possibly originating from the lack of mass selection in the MIS experiments. 

The large frequency range over which vibrational activity is observed is in line with the relatively low symmetry of the molecule (C$_{s}$). As a result, all twelve CH-stretch modes are formally IR active, although the harmonic calculations predict that only six of them have appreciable intensity (Fig.\ref{Fig3Benz}, top). The SP16 calculations, on the other hand, match the experimental spectrum very well. They predict activity of a plethora of vibrational transitions as is clear when the stick spectrum underlying the green trace in Fig. \ref{Fig3Benz} (middle panel) is inspected, and thereby confirm the conclusions that the the broad appearance of the spectrum is due to the overlapping of many anharmonic bands.

\subsection{Chrysene}

\begin{figure}[]
\includegraphics[scale=0.5]{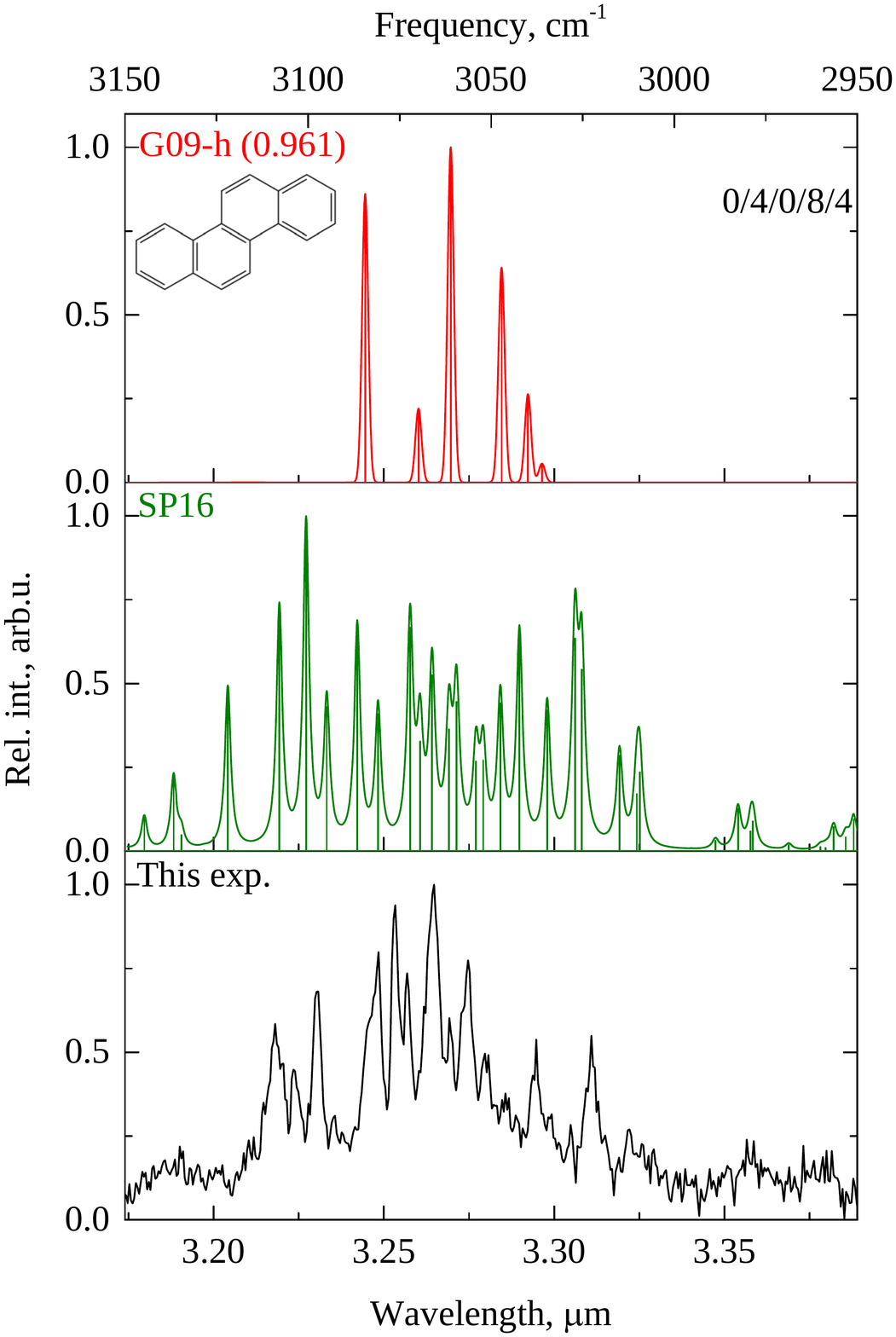}
\caption{The IR absorption spectrum of chrysene as predicted by G09-h and SP16 calculations together with the molecular beam gas-phase spectrum as measured in the present experiments.\label{Fig4(chrys)}}
\end{figure}

Chrysene (C$_{18}$H$_{12}$) is a highly symmetric isomer of benz[a]anthracene (C$_{2h}$  versus C$_{s}$). Its experimentally obtained and theoretically predicted IR absorption spectra are shown in Figure \ref{Fig4(chrys)}. In contrast to benz[a]anthracene, the experimental spectrum displays a larger number of resolved bands that have line widths which range from 3-7 cm$^{-1}$. Remarkably, major vibrational activity is found in the 3015-3115 cm$^{-1}$ range, which is even larger than for benz[a]anthracene, with the most intense band at 3063 cm$^{-1}$. In this respect, it is interesting to note the presence of bands above 3090 cm$^{-1}$ where no activity was observed for benz[a]anthracene. 

Table \ref{table2} reports the line positions of the fifteen bands that can be distinguished and compares them with previously reported MIS positions \citep{Hudgins1998b}. This comparison shows deviations less than 2 cm$^{-1}$ caused mainly by the inability to accurately identify line positions of close lying bands under the relatively low-resolution conditions of the MIS experiment. The gas-phase data furthermore identify eight previously unreported transitions, but does not show evidence for the MIS band at 3134.9 cm$^{-1}$. Symmetry considerations lead to the conclusion that at most six CH-stretch modes could be IR-active (Fig.\ref{Fig4(chrys)}, top panel) and that they are localized in a relatively narrow frequency range. SP16 calculations (Fig.\ref{Fig4(chrys)}, middle panel) once again emphasize the important role of anharmonicity and Fermi resonances, and give rise to a predicted spectrum that is in good agreement with the experiment.

\subsection{Triphenylene}

\begin{figure}[]
\includegraphics[scale=0.5]{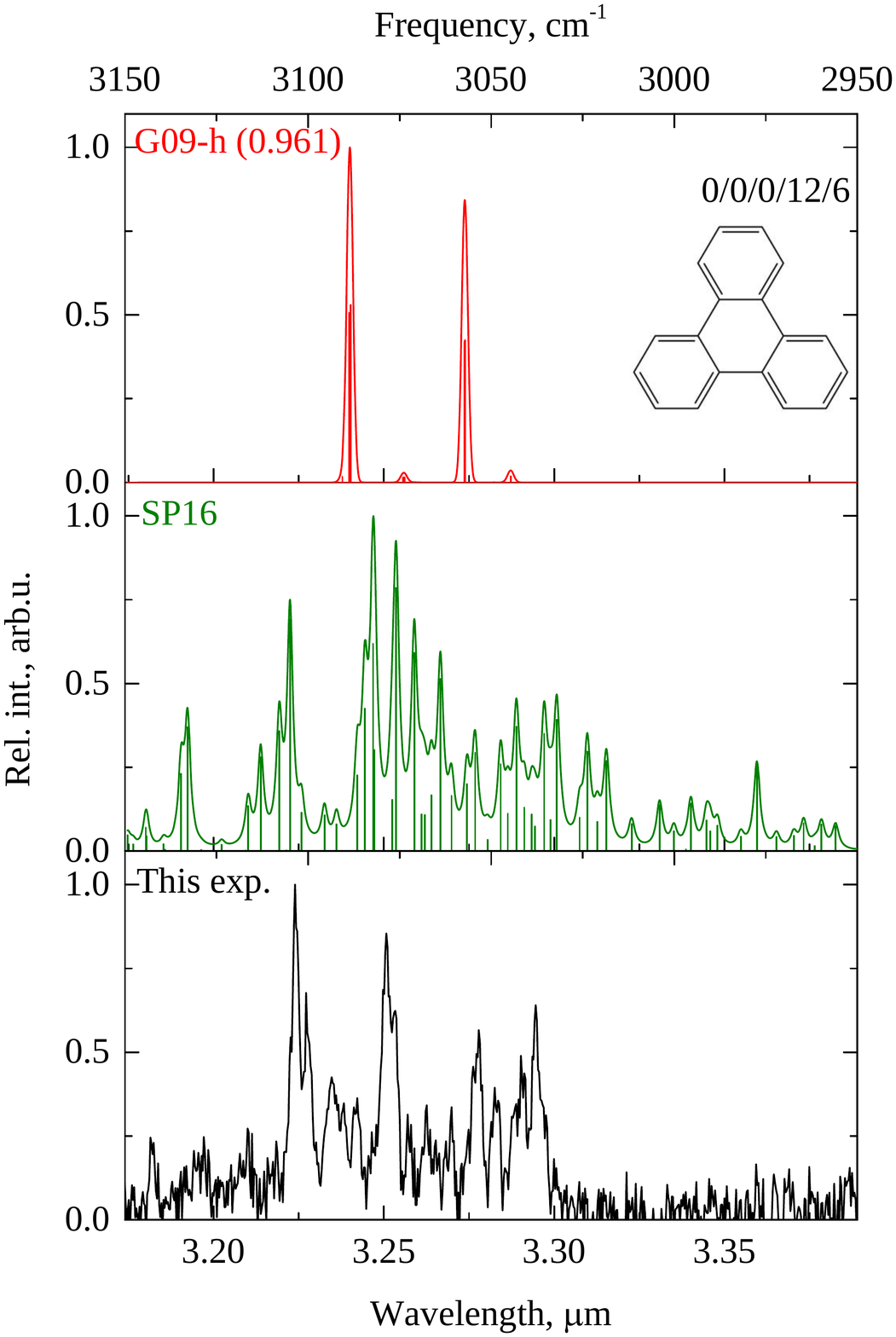}
\caption{The IR absorption spectrum of triphenylene as predicted by G09-h and SP16 calculations together with the molecular beam gas-phase spectrum as measured in the present experiments. \label{Fig5(triph)}}
\end{figure}

Triphenylene completes the series of PAHs with four aromatic rings. For this molecule no high-resolution gas-phase vibrational spectra have been reported previously. The 3-$\mu$m band recorded here using UV-IR ion dip spectroscopy is depicted in Figure \ref{Fig5(triph)} (bottom panel). Activity is observed over a relatively small range of 3030-3105 cm$^{-1}$ with three strong features at 3040, 3075 and 3100 cm$^{-1}$ and the strongest band at 3101.8 cm$^{-1}$. Above 3105 cm$^{-1}$, one band at 3143.2 cm$^{-1}$ appears to be present, there is no indication for bands below 3025 cm$^{-1}$. Overall, sixteen bands with an irregular shape and a minimum linewidth of 2.9 cm$^{-1}$ can be resolved. Considering their profile, it is likely that most of these bands actually consist of several overlapping transitions. Previous MIS studies \citep{Hudgins1998b} compare well with the present data (see Table \ref{table2}) with differences in band positions not exceeding 2 cm$^{-1}$. In the present study significantly more bands are observed than in the MIS studies, and the MIS 3116.9 cm$^{-1}$ band is not confirmed.

Triphenylene belongs to the D$_{3h}$ symmetry point group and because of this high symmetry only four pairs of doubly degenerate CH-stretch modes are IR active in the harmonic approximation. As Gaussian 09 can not perform the anharmonic analysis of molecules with D$_{3h}$ symmetry, the calculation was done on triphenylene with a reduced C$_{2v}$ symmetry caused by a small perturbation of the masses of two opposing carbon atoms, from 12 to 12.01 au (see \citet{Mackie2016} for further details).

Under such conditions a spectrum is predicted (Fig. \ref{Fig5(triph)}, top) with only two strong bands, which is clearly at odds with the experiment. SP16 calculations incorporating anharmonicity, on the other hand, redistribute the intensities among fundamentals and combination bands with equal symmetries and lead to a predicted absorption spectrum that resembles the experimentally observed one (Fig.\ref{Fig5(triph)}, middle panel). Nevertheless, a detailed comparison of the two spectra does lead to the conclusion that in the regions above 3105 and below 3025 cm$^{-1}$ quite a larger activity is predicted by the calculations than observed experimentally. In fact, similar observations can also be made for chrysene (regions above 3110 and below 3020 cm$^{-1}$) and to some extent for benz[a]anthracene (region above 3090 cm$^{-1}$) although in these cases the differences are not as obvious as in the case of triphenylene.

\subsection{Perylene}

\begin{figure}[]
\includegraphics[scale=0.5]{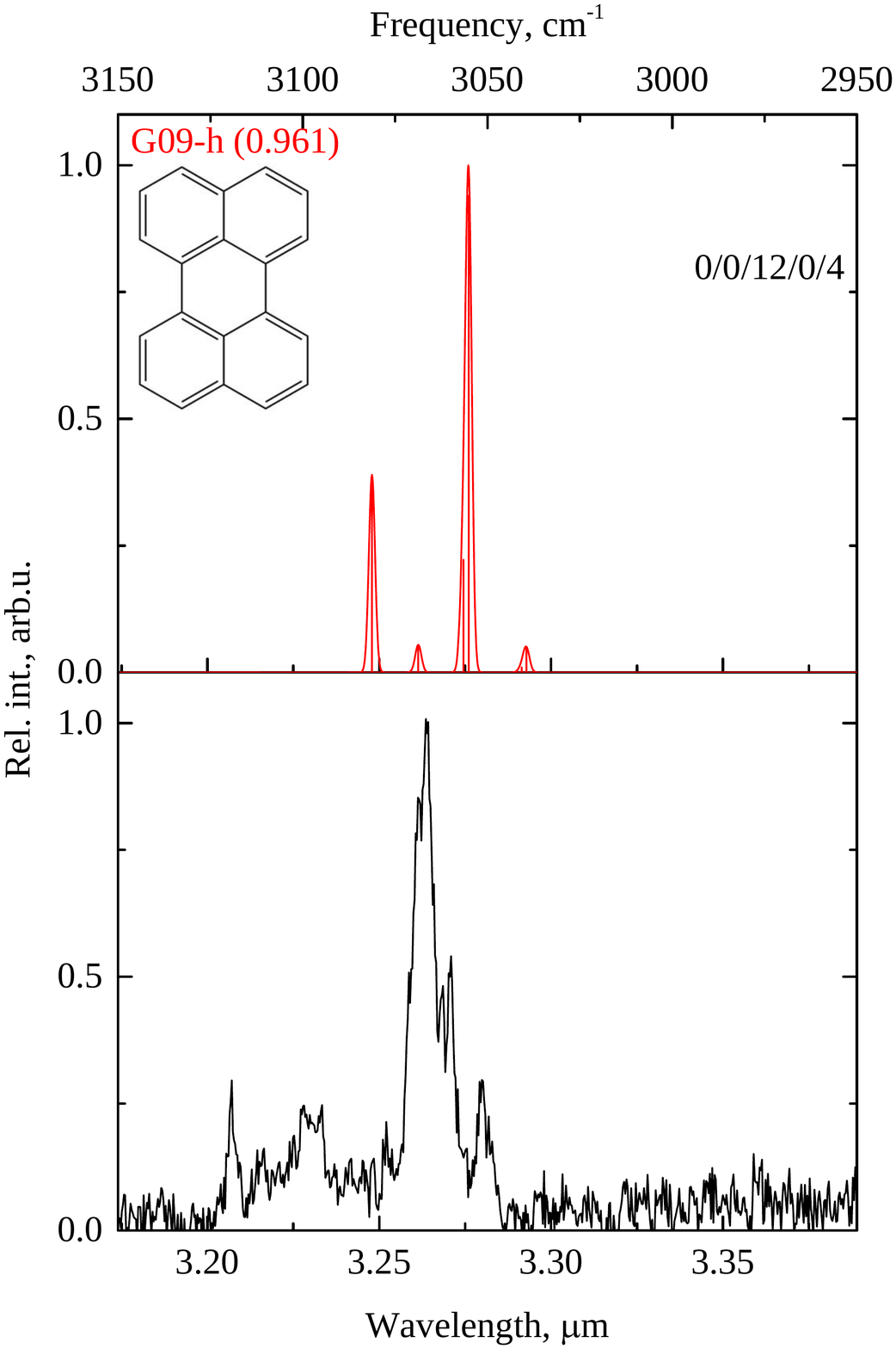}
\caption{The IR absorption spectrum of perylene as predicted by G09-h calculations together with the molecular beam gas-phase spectrum as measured in the present experiments. Calculations at the SP16 level lead to incorrect results (see text) and therefore are not reported.\label{Fig6(per)}}
\end{figure}

Perylene is the most condensed five-aromatic ring PAH system. The experimental IR absorption spectrum of this molecule in the 3-$\mu$m region is shown in the bottom panel of Figure \ref{Fig6(per)} together with predicted spectra at the G09-h level (top panel). The spectrum is quite compact and dominated by a group of bands located in 3056-3070 cm$^{-1}$ with the most intense transition occurring at 3063.8 cm$^{-1}$. At the high-energy side a weak, sharp band is observed at 3118 cm$^{-1}$ and a broad structure that appears to consist of three overlapping bands at 3092.8, 3095.9, 3098.2 cm$^{-1}$. Overall, nine bands with a linewidth of $\geqslant$3 cm$^{-1}$ are reported in Table \ref{table2}. Perylene has previously been studied with CRDS \citep{Huneycutt2004}, but the spectrum reported at that time clearly has a lower S/N ratio. This is most likely the reason that a comparison between the two spectra is not as favorable as for the other molecules. For example, we do not find indications for the previously reported bands at 3044.9 and 3070.4 cm$^{-1}$ and cannot confirm the band at 3022 cm$^{-1}$. Moreover, our cooling conditions enable us to conclude that the bands reported at 3096.1 cm$^{-1}$ and 3065.3 cm$^{-1}$ (CRDS) consist of three and two bands, respectively. 
Perylene possesses D$_{2h}$ symmetry and according to the harmonic approximation  only 6 CH-stretch modes are IR active. The other observed bands originate from anharmonic activities -- i.e., resonances. 

Unfortunately, it was not possible to perform an anharmonic analysis on perylene as  Gaussian09 was unable to produce reasonable cubic and quartic force constants. Those unreasonable force constants lead to a significant overestimation of the anharmonic corrections of band positions, with many bands predicted to occur at higher energies, which is contrary to what is expected. It would appear that these problems depend on the level of theory that is used since calculations on perylene using the B3LYP functional with a 4-31G basis set lead to cubic and quartic force constants that seem much more reasonable. However, this level of theory is too low to allow for meaningful comparisons with the experimental spectra. Further studies on the functional/basis set dependence of the quartic force field of large molecules such as perylene would thus be of interest, but fall outside the scope of the present studies. 

\section{Discussion}

 The high-resolution IR absorption spectra of the presently studied set of PAHs together with those of the linear acenes studied previously \citep{Maltseva2015} provide a comprehensive view on key factors that determine the appearance of the 3-$\mu$m band in PAHs. The number of CH-oscillators obviously controls the number of normal modes, and this clearly implies that compact species always have less active modes in the 3-$\mu$m region than more extended molecules of the same symmetry and with the same number of rings. However, from the comparison of 3-$\mu$m bands of PAHs with different molecular structure it becomes clear that the frequency region over which activity is observed and the distribution of intensity over this region is not as dependent on the number of CH-oscillators as one might expect.
 
\begin{figure}[]
\includegraphics[scale=0.5]{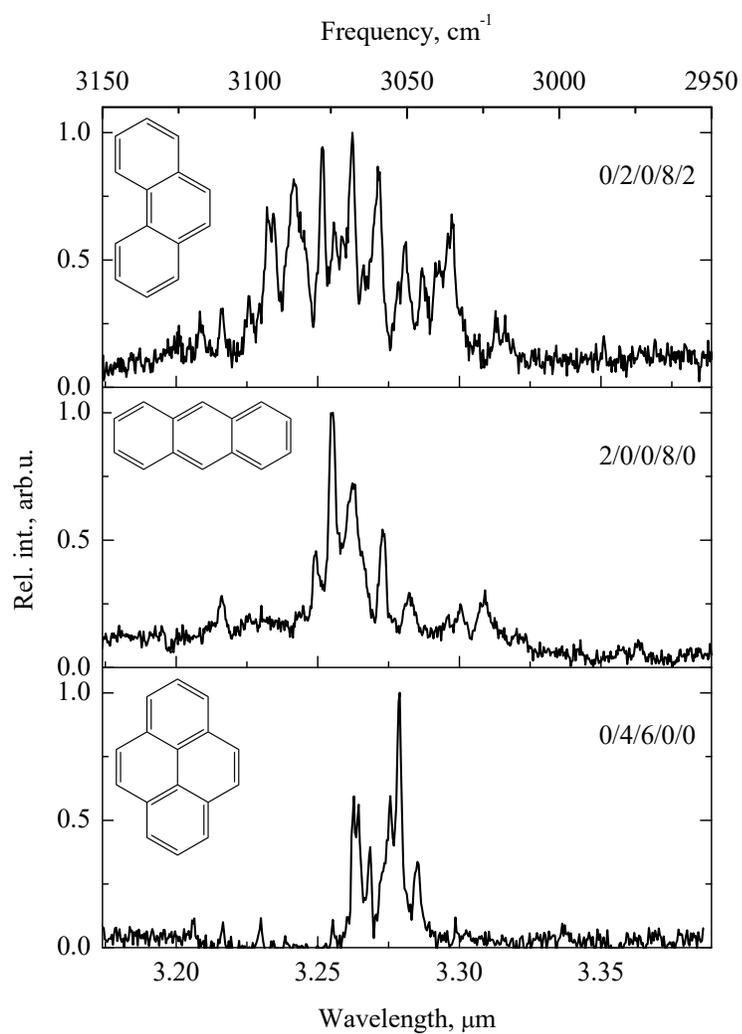}
\caption{Experimental IR absorption spectra of pyrene, anthracene and phenanthrene from the bottom to the top, respectively. Number of hydrogens is mentioned in the following order: solo/duo/trio/quartet/bay H's.\label{Fig7(Phen)}}
\end{figure}

\begin{figure}[]
\includegraphics[scale=0.5]{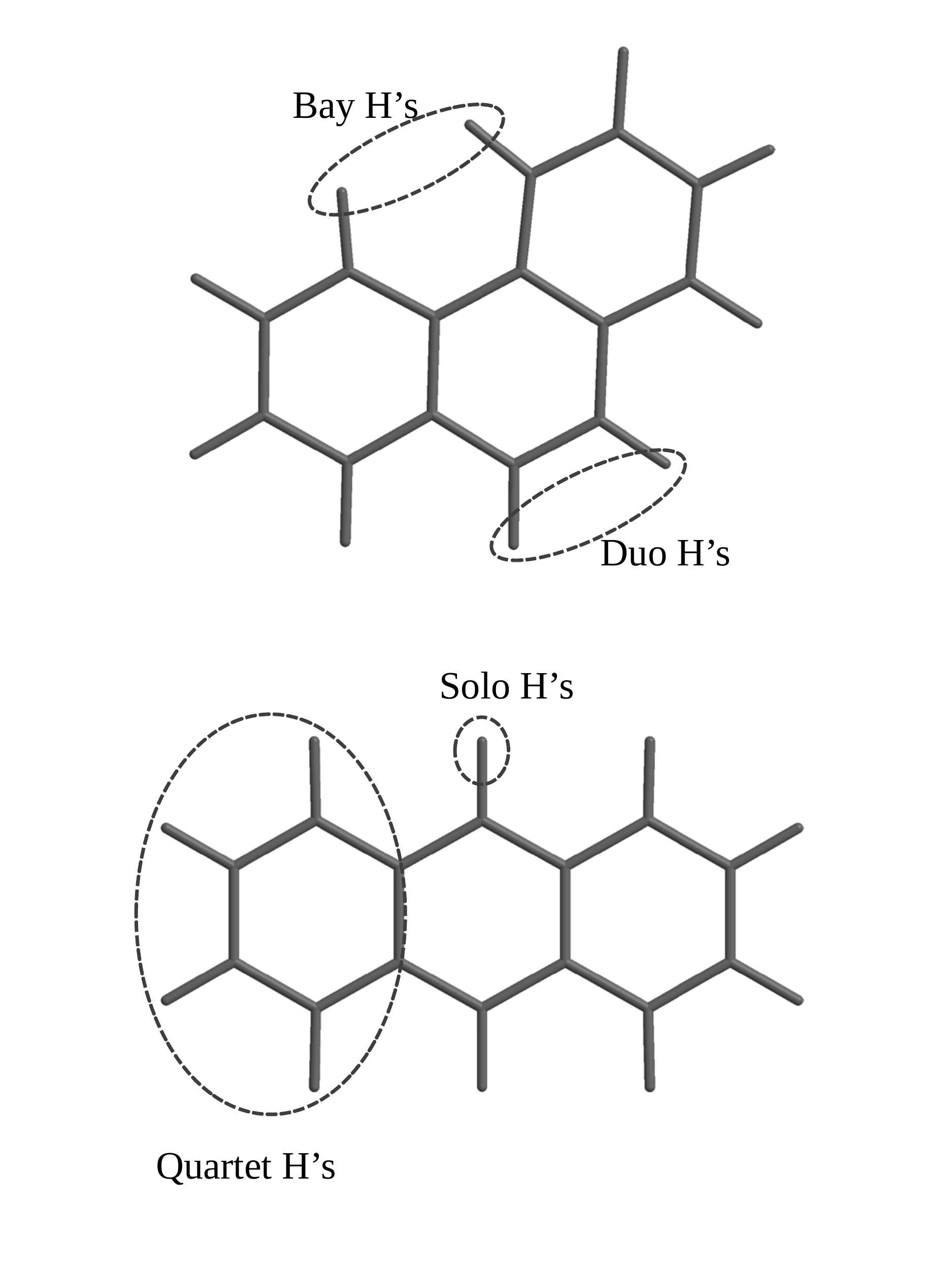}
\caption{Schematic representation of bay-region, solo, duo and quartet hydrogens using examples of anthracene (bottom) phenanthrene (top).\label{Fig8bay}}
\end{figure}

\begin{figure}[]
\includegraphics[scale=0.5]{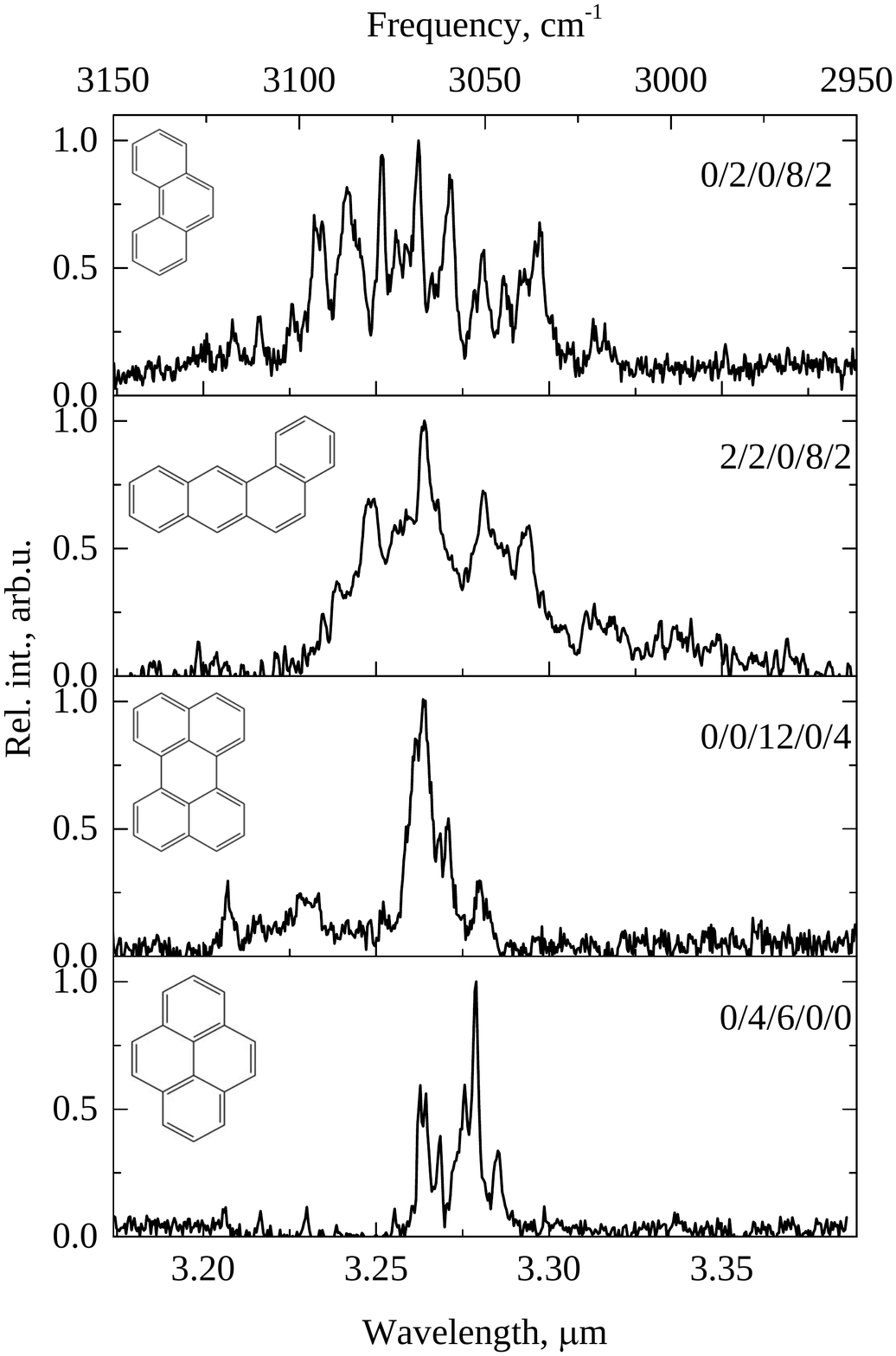}
\caption{Experimental IR absorption spectra of  pyrene, perylene, benz[a]anthracene and phenanthrene from the bottom to the top.\label{Fig10(perylene)}}
\end{figure}

 For example, four-ring pyrene consists of sixteen carbons and ten hydrogens and its 3-$\mu$m bands looks quite similar to anthracene (Fig.\ref{Fig7(Phen)}), another molecule with ten CH-oscillators and D$_{2h}$ symmetry. Both 3-$\mu$m bands are compact, the most active region occurring in a range of 22 and 30 cm$^{-1}$, respectively. However, a noticeable blue-shift of the strongest transition of anthracene (3071.9 cm$^{-1}$) with respect to pyrene (3049.8 cm$^{-1}$) is observed. 
Phenathrene is an isomer of anthracene, but despite just a small change in molecular structure, its 3-$\mu$m band is very different (Fig.\ref{Fig7(Phen)}). The most active region spans 60 cm$^{-1}$ with the most intense band at 3065.3 cm$^{-1}$. These differences likely originate from the nonlinear structure of phenanthrene. Significant changes occur in the periphery of the molecule. Anthracene has two solo-hydrogens and two quartets. Phenanthrene, on the other hand, has two duo hydrogens instead of two solos, and also two quartets, but conversely the quartets each have a hydrogen in the bay-region (Fig. \ref{Fig8bay}). As a result, steric effects increase the vibrational frequencies of these CH-oscillators \citep{Bauschlicher2009}. Interestingly, Figure  \ref{Fig10(perylene)} shows that the 3-$\mu$m bands of phenanthrene resembles to a large extent that of benz[a]anthracene. The most intense transitions of phenanthrene and  benz[a]anthracene are at 3065.3 and 3063.8 cm$^{-1}$, the most active regions are 3030-3090 and 3035-3097 cm$^{-1}$, respectively. This can be rationalized on the basis of their similar peripheral structure, the only difference between the two molecules being the additional two solo hydrogens in benz[a]anthracene, which do not contribute significantly to the spectrum.

The 3-$\mu$m absorption of pyrene looks very similar to the absorption of the five-ring PAH with twelve hydrogens - perylene (Fig.\ref{Fig10(perylene)}). The most active modes of perylene are concentrated in a narrow frequency region of 15 cm$^{-1}$. This relatively small range follows quite nicely from the molecular and, in particular, the peripheral structure of perylene. All hydrogens in perylene come as trios and thus contribute to IR absorption at similar frequencies. Perylene and pyrene indeed have a similar band at 3064 cm$^{-1}$ which is attributed to modes localized in both cases in trio hydrogens. Pyrene does not possess bay-hydrogens, but perylene does. As confirmed by the calculations this explains the differences in activity in the high-frequency regions of the spectra of the two compounds. 

 From the subsequent discussion (vide infra) it appears that   molecules with the same type of hydrogens (solos, duos, trios, or quartets) display activity in a similar frequency range. Moreover, from the comparison of our experiment with DFT theory,  we suggest that the frequencies of normal modes that primarily involve motion of the same type of hydrogens are restricted  to rather narrow frequency regions with frequencies that increase in the order of solos, duos, trios, to quartet \citep{Mackie2016}.

\begin{figure}[]
\includegraphics[scale=0.5]{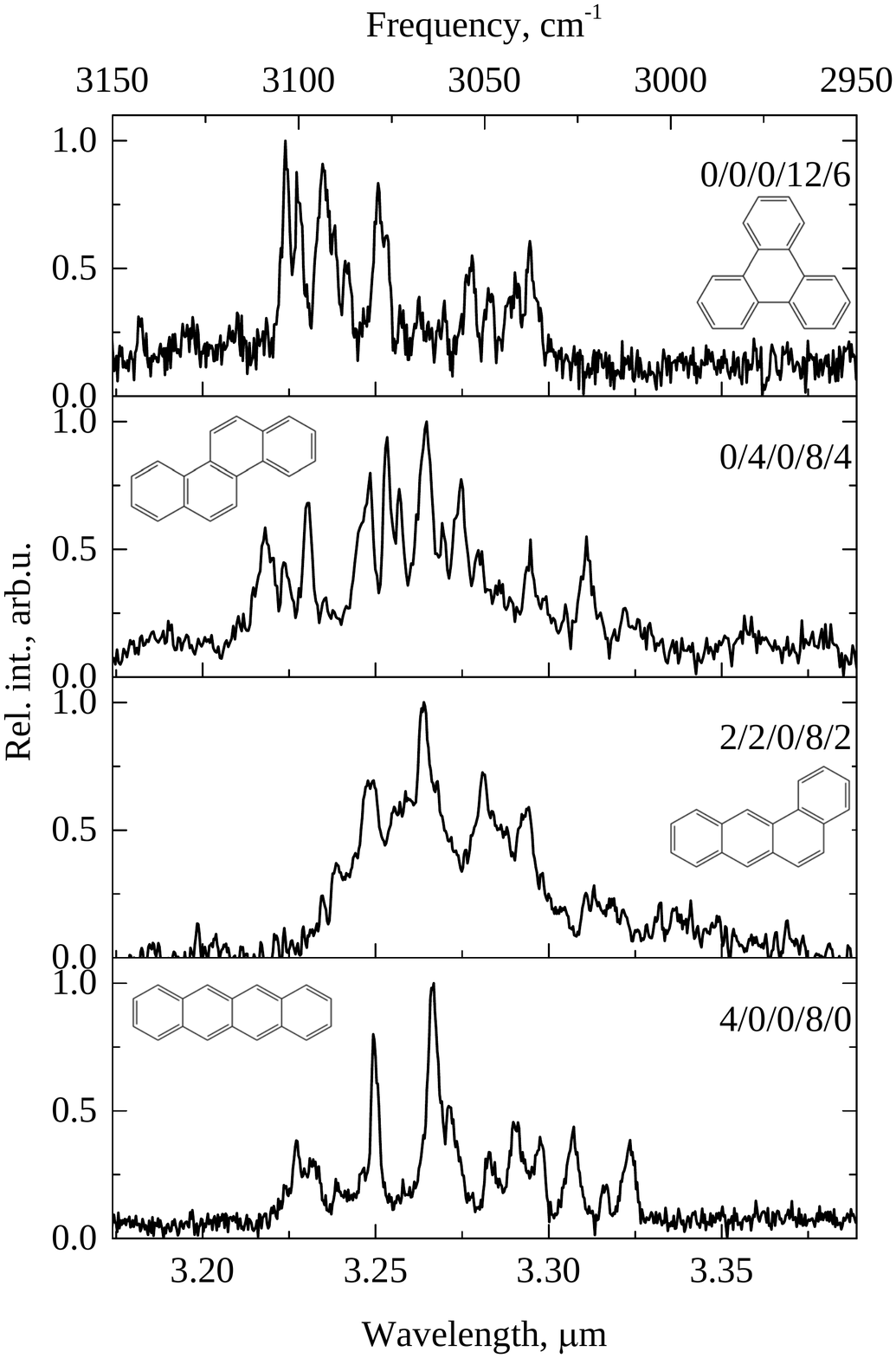}
\caption{Experimental IR absorption spectra of  tetracene, benz[a]antracene, chrysene, triphenylene from the bottom to the top.\label{Fig9(isomers)}}
\end{figure}

Another effect induced by steric hindrance associated with bay regions is best illustrated by comparing the 3-$\mu$m absorption of tetracene with its isomers in Fig.\ref{Fig9(isomers)}. Benz[a]anthracene has a pair of hydrogens (one is from a quartet and another one is a solo hydrogen) in a bay-site. Its 3-$\mu$m band has a maximum at 3063.8 cm$^{-1}$, close to the most intense transition of tetracene (3061.1 cm$^{-1}$) and major activity over 3035-3097 cm$^{-1}$. While the strongest transition of chrysene (3063 cm$^{-1}$) is not affected by the structural changes and remains very close to tetracene, the high-energy part of the spectrum   
undergoes changes. Chrysene shows prominent high-energy features in the 3090-3120 cm$^{-1}$ region where tetracene has considerably less activity (Fig.\ref{Fig9(isomers)}). We conclude that this activity should be attributed to bands involving modes with considerable contributions of hydrogens in the bay region as these have intrinsically higher vibrational frequencies because of steric hindrance. The conclusion is supported by the SP16 calculations. It is found computationally that all resonances in this region involve the CH-stretch mode of chrysene located in a bay region. All hydrogens of triphenylene are incorporated into three quartets with half of them in bay-sites. In contrast to the other PAHs with quartets, triphenylene has the most intense band at 3101.8 cm$^{-1}$ while for the other compounds it is found between 3063-3066 cm$^{-1}$ (Fig.\ref{Fig9(isomers)}). This is a clear demonstration of the presence of bay-site hydrogens. The band at 3076.2 cm$^{-1}$ most likely dominantly involves asymmetrically stretching of bay-hydrogens. The 3030-3060 cm$^{-1}$ region can nicely be explained by anharmonic activity \citep{Mackie2016}. Such a comparison clearly shows that vibrations involving hydrogens in bay-site are blue-shifted. 
 
 The symmetry of the molecule controls the number of IR-allowed transitions. These transitions are the source of intensity for combination bands of the same symmetry that normally would be 'dark' but can acquire intensity through Fermi resonance. When comparing isomers with different symmetries, for example anthracene and phehanthrene, the isomer with a higher symmetry (anthracene) is thus expected to show a smaller number of fundamental and combination bands as is indeed confirmed by our experiments (Fig.\ref{Fig7(Phen)}). Therefore, molecules with low degree of symmetry, like benz[a]anthracene, demonstrate broad absorption because of the overlap of anharmonic bands (Fig.\ref{Fig3Benz}).

 Another important conclusion that can be drawn from the present experiments concerns the frequency range over which anharmonicity is active in the 3-$\mu$m region. We have acquired high-resolution IR absorption spectra for PAHs with two to five rings in all possible geometrical arrangements, but for all these species vibrational activity quickly disappears below, roughly speaking, 3000 cm$^{-1}$ (above 3.33 $\mu$m). This conclusion is supported by our SP16 calculations that also show very limited activity in this region. Moreover,  in the SP16 calculations an important parameter that controls the number of states to be included in resonanaces/polyads is the maximum energy separation of the states. In all our calculations we find that a maximum separation of 200 cm$^{-1}$ is more than adequate. On the basis of the scaled harmonic calculations in which the lowest frequencies occur around 3025 cm$^{-1}$  one can then conclude that the maximum wavelength range over which combination band activity might be observed extends only up to 2825 cm$^{-1}$ (3.54 $\mu$m), and this number is most likely even an overestimate. Increasing the size of a PAH will thus increase the density of `dark' states that might potentially couple to the `bright' zeroth-order states, but the present study strongly suggests that this increased density has a very limited effect on the range of the region over which vibrational activity is observed.

\section{Astrophysical implications}

The present study confirms the dominant role of anharmonicity in determining the shape and strength of the features in the CH-stretching region. Predictions by harmonic calculations thus are far from adequate for obtaining a fundamental understanding of how the 3-$\mu$m region reflects the chemical composition and evolution of astronomical objects, and for trying to understand secondary features around the 3-$\mu$m feature. 
They also indicate that in principle there are many more leads available for identifying single PAH species as intensity is distributed over many more transitions than only the fundamental CH-stretch transitions.

One of the secondary features that has been discussed extensively in literature concerns the two-component emission character of the 3-$\mu$m band \citep{Song2003a}. Our results clearly show that bay-hydrogens induce intensity at the high-energy side, and thus fully support the previous suggestion that the 3.28-$\mu$m emission band is associated with modes involving bay-hydrogens as occurring in more extended PAHs, while the 3.30-$\mu$m emission band derives from more compact PAH structures \citep{Candian2012}. The size of the PAH, which also has been put forward as the primary cause for the two-component emission \citep{Song2007}, appears to be only of secondary influence. Nevertheless, our results also unmistakably demonstrate that activity in the high-frequency range is not only induced by bay-hydrogens, but also occurs for PAHs without bay-sites. The finer details of this part of the spectrum thus are the result from a subtle interplay between the effects of steric hindrance as encountered for the bay hydrogens and anharmonicities. This implies that one should be cautious in directly relating  the ratio of the intensities of the two bands to abundances of compact and extended PAH structures.

Another secondary feature that has attracted considerable attention is the 3-$\mu$m plateau spanning the 3.1-3.7 $\mu$m (2700-3200 cm$^{-1}$) region \citep{Allamandola1989a,Geballe1989}. Our spectra show that the intensity in the high-energy part of this plateau is derived from Fermi resonances between fundamental CH-stretch transitions and the plethora of combination bands that are present in this region. However, such an explanation, which previously has also been put forward as a possible cause for the lower-energy side of the plateau \citep{Allamandola1989a}, is not supported by the present results which show only very limited activity in the region above 3.3 $\mu$m (below 3000 cm$^{-1}$). Above, we have reasoned that also for larger PAHs we do not expect an increased activity in this region. We therefore conclude that in this region IR absorption has another origin.

A final feature of interest is the 3.40-$\mu$m feature on the 3-$\mu$m plateau. Several explanations have been put forward to account for its presence ranging from hot bands of aromatic CH-stretch transition $\upsilon$ = 2 $\rightarrow$ 1 shifted due to anharmonic effects \citep{Barker1987} to CH-stretch modes in methylated \citep{Joblin1996,Pauzat1999} and superhydrogenated \citep{Bernstein1996,Steglich2013,Sandford2013,Wagner2000} PAHs
, but none of these have so far been confirmed. The overall picture that emanates from our high-resolution studies, involving in particular the role of anharmonicity, suggests that an explanation based on hydrogenated and alkylated PAHs is attractive. Methylated and hydrogenated PAHs have been known to be IR active at much lower frequencies than the fully aromatic systems and might thus contribute to the 3-$\mu$m plateau. Assuming that anharmonicity plays a similar role in such alkylated PAHs, this would then also explain the extended range of the 3-$\mu$m plateau which cannot be explained merely in terms of bare PAHs. To find further support for such a conclusion we are presently performing high-resolution IR absorption studies on an extensive series of hydrogenated and alkylated PAHs.

\section{Conclusions}

In this work we have presented molecular beam IR absorption spectra of six condensed PAHs in the 3-$\mu$m region using IR-UV ion dip spectroscopy. The present results and the results on linear acenes \citep{Maltseva2015, Mackie2015} show that anharmonicity indeed rules the 3-$\mu$m region, with the fraction of intensity not associated with fundamental transitions easily exceeding 50$\%$. Anharmonicity-induced transitions are more the rule than the exception and should explicitly be taken into account in the interpretation of astronomical data in the 3-$\mu$m region. A proper incorporation of resonances has been shown to yield predicted spectra that are in semi-quantitative agreement with the experiments. Such calculations may therefore lead the way for furthering our understanding on the influence of larger PAHs which are not amenable to similar experimental high-resolution studies.

Our work shows that the observed abundance of combination bands is mostly concentrated in the low-energy part ($\leq$ 3100 cm$^{-1}$) of the CH-stretch region and originates from combinations of CC-stretch and CH in-plane bending modes. This anharmonic activity can be partly responsible for the 3-$\mu$m plateau observed by astronomers although both experiment and theory put into question a scenario in which the plateau is solely attributed to Fermi resonances between fundamental modes and such combination bands. In this respect, a more important role than assumed so far of hydrogenated and alkylated PAHs in combination with anharmonic effects appears to provide a highly-interesting alternative to pursue. Activity at the high-energy part of the 3-$\mu$m band has been demonstrated to derive from the presence of bay-hydrogens, but not solely as also anharmonicity in PAHs without bay-hydrogens induces activity in this region. Further studies on larger compact PAHs should provide further means to distinguish between the relative importance of the two effects. Such studies are presently underway.

Our studies show that the vibrational activity in the CH-stretch fingerprint region is strongly linked to details of the molecular structure. It is therefore not only different for molecules with different chemical structures, but also for different isomers. We have demonstrated that the peripheral structure of the molecules plays a dominant role in the appearance of the 3-$\mu$m band. Studies in which the 3-$\mu$m region is correlated with the "periphery-sensitive" 9-15 $\mu$m region are thus of significant interest and could further elucidate the exact composition of the carriers of these bands. 
\\

The experimental work was supported by The Netherlands Organization for Scientific Research (NWO). AP acknowledges NWO for a VIDI grant (723.014.007). Studies of interstellar PAHs at Leiden Observatory have been supported through the advanced European Research Council Grant 246976 and a Spinoza award. Computing time has been made available by NWO Exacte Wetenschappen (project MP-270-13 and MP-264-14) and calculations were performed at the LISA Linux cluster and Cartesius supercomputer (SurfSARA, Almere, NL). AC acknowledges NWO for a VENI grant (639.041.543). XH and TJL gratefully acknowledge support from the NASA 12-APRA12-0107 grant. XH acknowledges the support from NASA/SETI Co-op Agreement NNX15AF45A. Some of this material is based upon work supported by the National Aeronautics and Space Administration through the NASA Astrobiology Institute under Cooperative Agreement Notice NNH13ZDA017C issued through the Science Mission Directorate

\onecolumn
\begin{table}
\begin{center}
\caption{Frequencies (cm$^{-1}$) and intensities of phenanthrene, pyrene and benz[a]antracene obtained in our study and compared to the data from cavity ring down spectroscopy (CRDS)\citep{Huneycutt2004} and matrix-isolation spectroscopy (MIS) data \citep{Hudgins1998b}.\label{table1}}
\end{center}
\begin{tabular}{llrllrllr}

\tableline\tableline
\multicolumn{3}{c}{Phenanthrene (C$_{14}$H$_{10}$)}&\multicolumn{3}{c}{Pyrene (C$_{16}$H$_{10}$)}&\multicolumn{3}{c}{benz[a]antracenee (C$_{18}$H$_{12}$)}\\

\multicolumn{2}{c}{this work} & CRDS& \multicolumn{2}{c}{this work}	& CRDS & \multicolumn{2}{c}{this work}& MIS				\\

\cline{1-2} \cline{4-5} \cline{7-8}

freq.& rel. int.& freq.&freq.& rel. int.& freq.&freq.& rel. int.& freq.\\

\tableline

3015.5 & 0.25 &        & 3044   & 0.34 & 3045.9 & 3017.1 & 0.2  & 3016.5 \\
3018.6 & 0.26 & 3018.7 & 3049.8 & 1    & 3051.7 & 3036.8 & 0.57 &        \\
3032.8 & 0.68 & 3034.7 & 3052.9 & 0.6  &        & 3047.8 & 0.73 & 3044.5 \\
3037.6 & 0.47 & 3038.5 & 3055.6 & 0.29 &        & 3063.8 & 1    & 3064.1 \\
3042.3 & 0.45 &        & 3057.5 & 0.13 &        & 3079   & 0.69 & 3077.6 \\
3042.4 & 0.45 & 3044.3 & 3059.5 & 0.39 & 3058.1 & 3087.8 & 0.37 &        \\
3047.9 & 0.55 &        & 3063.3 & 0.56 & 3061.3 &        &      &        \\
3050.3 & 0.4  & 3049.1 & 3064.9 & 0.59 & 3064.5 &        &      &        \\
3056.4 & 0.88 & 3057.3 & 3067.1 & 0.12 & 3067.3 &        &      &        \\
3057   & 0.9  & 3058.5 & 3071.9 & 0.11 & 3071.7 &        &      &        \\
3061.8 & 0.5  &        & 3087.8 & 0.08 & 3090   &        &      &        \\
3065.3 & 1    & 3066.9 & 3096   & 0.12 & 3094.8 &        &      &        \\
3069   & 0.57 &        & 3108.9 & 0.1  & 3108.4 &        &      &        \\
3071.6 & 0.64 & 3072.3 & 3118.7 & 0.12 &        &        &      &        \\
3075.4 & 0.89 & 3076.7 &        &      &        &        &      &        \\
3081.5 & 0.5  &        &        &      &        &        &      &        \\
3082.8 & 0.53 &        &        &      &        &        &      &        \\
3084.3 & 0.55 & 3083.9 &        &      &        &        &      &        \\
3091.8 & 0.3  &        &        &      &        &        &      &        \\
3094   & 0.33 & 3093.7 &        &      &        &        &      &        \\
3100   & 0.28 & 3101.3 &        &      &        &        &      &        \\
3109.6 & 0.25 & 3109.7 &        &      &        &        &      &        \\
3116.6 & 0.24 &        &        &      &        &        &      &       
\\

\end{tabular}
\end{table}

\begin{table}
\begin{center}
\caption{Frequencies (cm$^{-1}$) and intensities of chrysene, triphenyene and perylene obtained in our study and compared to the data from cavity ring down spectroscopy (CRDS)\citep{Huneycutt2004} and matrix-isolation spectroscopy (MIS) data \citep{Hudgins1998b}.\label{table2}}
\end{center}
\begin{tabular}{llrllrllr}

\tableline\tableline
\multicolumn{3}{c}{Chrysene (C$_{18}$H$_{12}$)}&\multicolumn{3}{c}{Triphenylene (C$_{18}$H$_{12}$)}&\multicolumn{3}{c}{Perylene (C$_{18}$H$_{12}$)}\\

\multicolumn{2}{c}{this work} & MIS& \multicolumn{2}{c}{this work}	& MIS & \multicolumn{2}{c}{this work}& CRDS				\\

\cline{1-2} \cline{4-5} \cline{7-8}

freq.& rel. int.& freq.&freq.& rel. int.& freq.&freq.& rel. int.& freq.\\

\tableline

3010.1 & 0.2  &        & 3035.3 & 0.64 &        & 3048.9 & 0.3  & 3050.3 \\
3020.3 & 0.5  & 3021.9 & 3039.4 & 0.5  & 3038.8 & 3057.4 & 0.5  &        \\
3035.1 & 0.5  & 3033.4 & 3040.6 & 0.35 &        & 3059.5 & 0.48 & 3060.9 \\
3043.5 & 0.32 &        & 3046.4 & 0.4  &        & 3063.8 & 1    & 3065.3 \\
3049   & 0.45 &        & 3050.8 & 0.57 & 3049.4 & 3066.3 & 0.85 &        \\
3053.7 & 0.75 & 3054.1 & 3058.2 & 0.34 &        & 3092.8 & 0.24 &        \\
3058.8 & 0.56 &        & 3065   & 0.34 &        & 3095.9 & 0.21 & 3096.1 \\
3063   & 1    & 3063.9 & 3070.3 & 0.31 &        & 3098.2 & 0.24 &        \\
3070.5 & 0.71 &        & 3073.8 & 0.62 & 3074.7 & 3118   & 0.3  &        \\
3073.8 & 0.93 &        & 3076.2 & 0.85 &        &        &      &        \\
3078.4 & 0.78 & 3079.4 & 3084.3 & 0.36 &        &        &      &        \\
3091   & 0.23 &        & 3088.1 & 0.35 &        &        &      &        \\
3095.7 & 0.65 & 3093.8 & 3091.5 & 0.43 & 3090.9 &        &      &        \\
3102.3 & 0.4  &        & 3098.7 & 0.68 & 3098   &        &      &        \\
3107.4 & 0.54 & 3106.8 & 3101.8 & 1    &        &        &      &        \\
       &      &        & 3143.2 & 0.2  &        &        &      &       
\end{tabular}
\end{table}

\end{document}